\documentclass[nohyper,11pt,letterpaper]{JHEP}
\usepackage[dvips]{epsfig}
\usepackage{epsf,amsfonts,amssymb}

\newcommand{\eqn}[1]{(\ref{#1})}
\newcommand{\be}{\begin{equation}}
\newcommand{\ee}{\end{equation}}
\newcommand{\ben}{\begin{displaymath}}
\newcommand{\een}{\end{displaymath}}
\newcommand{\bea}{\begin{eqnarray}}
\newcommand{\eea}{\end{eqnarray}}
\newcommand{\bean}{\begin{eqnarray*}}
\newcommand{\eean}{\end{eqnarray*}}
\newcommand{\nn}{\nonumber \\}
\newcommand{\ba}{\begin{array}}
\newcommand{\ea}{\end{array}}
\newcommand{\bi}{\begin{itemize}}
\newcommand{\ei}{\end{itemize}}

\def\a {\alpha}

\def\g {\gamma}
\def\G {\Gamma}
\def\d {\delta}
\def\s {\sigma}
\def\e {\epsilon}

\renewcommand{\t}{\theta}

\newcommand{\calb}{\mbox{${\cal B}$}}
\newcommand{\calc}{\mbox{${\cal C}$}}

\newcommand{\cale}{\mbox{${\cal E}$}}

\newcommand{\caln}{\mbox{${\cal N}$}}

\newcommand{\bx}{\mbox{\boldmath $x$}}
\newcommand{\by}{\mbox{\boldmath $y$}}
\newcommand{\bX}{\mbox{\boldmath $X$}}
\newcommand{\bY}{\mbox{\boldmath $Y$}}

\newcommand{\bbe}[1]{\mbox{${\mathbb E}^{#1}$}}
\newcommand{\bbr}[1]{\mbox{${\mathbb R}^{#1}$}}

\newcommand{\nef}{{$\caln =4$} }

\newcommand{\dd}[3]{\mbox{$( #1 | \mbox{D} #2 \perp \mbox{D} #3)$}}

\newcommand{\ads}[1]{\mbox{${AdS}_{#1}$}}
\newcommand{\adss}[2]{\mbox{$AdS_{#1}\times {S}^{#2}$}}

\newcommand{\pa}{\partial}
\newcommand{\fc}{\frac}
\newcommand{\w}{\wedge}

\newcommand{\sac}{\, , \qquad}

\newcommand{\sgn}[1]{\mbox{sgn}(#1)}

\newcommand{\ifive}{{\it 5}}

\newcommand{\ra}{\rightarrow}

\newcommand{\sect}[1]{\setcounter{equation}{0}\section{#1}}
\renewcommand{\theequation}{\arabic{section}.\arabic{equation}}

\title{\Large Supersymmetric Defect Expansion in CFT 
from AdS Supertubes}

\author{David Mateos, Selena Ng and Paul K.\ Townsend \\
   Department of Applied Mathematics and Theoretical Physics\\
   Centre for Mathematical Sciences \\
   Wilberforce Road, Cambridge CB3 0WA, United Kingdom \\
E-mail: \email{D.Mateos, S.K.L.Ng, P.K.Townsend@damtp.cam.ac.uk}}

\abstract{The AdS/dCFT correspondence is used to show that a
planar $q$-dimensional superconformal CFT defect expands, 
under the addition of electric charge and angular momentum, to a
supersymmetric higher-dimensional defect of geometry 
$\bbr{q} \times \calc$, where $\calc$ is an {\it arbitrary} 
curve. The dual string theory process is the expansion of
D-branes and fundamental strings into a supertube in an \ads{}
background. }


\preprint{DAMTP-2002-88 \\ \tt{hep-th/0207136}}

\begin{document}

%%%%%%%%%%%%%%%%%%%%%%%%%%%%%%%%%%%%%%%%%%%%%%%%%%%%%%%%%%%%%%%%%%%%%%%%%%%%%%
%%%%%%%%%%%%%%%%%%%%%%%%%%%%%%%%%%%%%%%%%%%%%%%%%%%%%%%%%%%%%%%%%%%%%%%%%%%%%%
%%%%%%%%%%%%%%%%%%%%%% SECTION 1 %%%%%%%%%%%%%%%%%%%%%%%%%%%%%%%%%%%%%%%%%%%%%
%%%%%%%%%%%%%%%%%%%%%%%%%%%%%%%%%%%%%%%%%%%%%%%%%%%%%%%%%%%%%%%%%%%%%%%%%%%%%%
%%%%%%%%%%%%%%%%%%%%%%%%%%%%%%%%%%%%%%%%%%%%%%%%%%%%%%%%%%%%%%%%%%%%%%%%%%%%%%
%\newpage
\sect{Introduction}
%%%%%%%%%%%%%%%%%%%%%%%%%%%%%%%%%%%%%%%%%%%%%%%%%%%%%%%%%%%%%%%%%%%%%%%%%%%%%%
%%%%%%%%%%%%%%%%%%%%%%%%%%%%%%%%%%%%%%%%%%%%%%%%%%%%%%%%%%%%%%%%%%%%%%%%%%%%%%

The AdS/dCFT correspondence \cite{KR01} is an interesting extension of
the AdS/CFT correspondence \cite{Maldacena97} in which additional
structure is added on both sides. On the string theory side a brane
that extends to the \ads{} boundary is added. On the CFT side we have a
defect where the brane meets the \ads{} boundary. A necessary
condition for the defect to preserve (part of) the conformal
invariance of the bulk CFT is that the induced metric on the brane
include an \ads{} factor; we will see, however, that this is not
sufficient. 

In type IIB string theory, an AdS/dCFT correspondence can be
`derived' for each orthogonal $q$-dimensional intersection
of $N$  D3-branes and one D$p$-brane, which we shall denote as
\dd{q}{3}{p}. For example, consider the 1/4-supersymmetric 
\dd{2}{3}{5} intersection represented by the array
\be
\begin{array}{rcccccccccl}
\mbox{D3:}\,\,\, & 1 & 2 & 3 & \_ & \_ & \_ & \_ & \_ & \_ & \, \\
\mbox{D5:}\,\,\, & \_ & 2 & 3 & 4 & 5 & 6 & \_ & \_ & \_ & .
\ea
\label{d3/d5}
\ee
This is the case originally introduced in \cite{KR01}. If $g_s \ll 1$ and $g_s N \gg 1$,
where $g_s$ is the string coupling constant, then the perturbative 
description of this system in the decoupling limit is
in terms of an \adss{4}{2} D5-brane probe in the \adss{5}{5}
near-horizon geometry of the D3-branes; the probe intersects the \ads{}
boundary at an \mbox{$\bbr{2}$-defect}. The \ads{4}-factor arises from the
\mbox{023-directions} together with the radial coordinate in the
\mbox{456-space}, whereas the two angular coordinates in this space yield
the $S^2$-factor. On the other hand, if $g_s N \ll 1$
then the perturbative description of the system is that of the \nef 
super Yang-Mills (SYM) theory in the presence of an $\bbr{2}$-defect. 
The AdS/dCFT correspondence states that these two descriptions are
actually equivalent throughout the entire parameter space.
The degrees of freedom on the defect arise from the the open strings 
connecting the D5-brane to the $N$ D3-branes \cite{Sethi97, KS98}, 
and are the holographic duals of the open-string degrees of freedom 
on the probe D5-brane \cite{DFO01}. 

More generally, every 1/4-supersymmetric \dd{q}{3}{p} intersection leads
to an analogous duality involving a probe \adss{q+2}{p-q-2} D$p$-brane in
\adss{5}{5} and an $\bbr{q}$-defect in the SYM theory. 
Since each probe D$p$-brane preserves eight Poincar\'e and eight
special conformal supersymmetries \cite{ST02} and (part of) the
conformal symmetry in \adss{5}{5}, so does the dual defect in the 
gauge theory. 

In the IIB Minkowski vacuum, the addition of angular momentum to a 
collection of orthogonal D$p$-branes and fundamental
strings causes them to expand to a D$(p+2)$-brane supertube 
\cite{MT01}, which we henceforth refer to as a D$(p+2)$-tube. This is a
1/4-supersymmetric tubular D$(p+2)$-brane of geometry 
$\bbr{p+1} \times \calc$, where the cross-section $\calc$ 
is a completely {\it arbitrary} curve\footnote{
Although the expansion always leads to a closed cross-section because
of charge conservation, \mbox{1/4-supersymmetry} allows it to be an
`open' curve without boundary (which must therefore extend to
infinity).}  
in \bbr{8-p} \cite{MNT01}; the initial strings and D$p$-branes appear
on the D$(p+2)$-tube as dissolved charges represented by electric and
magnetic Born-Infeld (BI) fields. 
Although originally found as a solution of the Dirac-Born-Infeld
(DBI) action, the supertube is an exact solution (to all orders in $\a'$)
of classical open-string theory \cite{KMPW02}. 
One crucial feature of the D$(p+2)$-tube is that it preserves the 
same supersymmetries as a collection of strings and D$p$-branes; 
in particular, there is no trace of a condition
associated to the presence of the D$(p+2)$-brane \cite{MT01}.
This paper is based on the observation that the supersymmetric
expansion to a D$(p+2)$-tube just described also takes place 
in the presence of (suitably oriented) D3-branes, in particular in their
\adss{5}{5} background. The holographically dual process in the CFT is
a {\it supersymmetric defect expansion}. 

Consider for concreteness the 1/8-supersymmetric intersection
\be
\begin{array}{rcccccccccl}
\mbox{D3:}\,\,\, & 1 & 2 & 3 & \_ & \_ & \_ & \_ & \_ & \_ & \, \\
\mbox{D3:}\,\,\, & \_ & \_ & 3 & 4 & 5 & \_ & \_ & \_ & \_ & \, \\
\mbox{F1:}\,\,\, & \_ & \_ & \_ & \_ & \_ & 6 & \_ & \_ & \_ & .
\ea
\label{d3-d3-f1}
\ee
This can be viewed as a \dd{1}{3}{3} intersection to which fundamental
strings have been added. 
If the D3-branes in the first line are replaced by an \adss{5}{5}
geometry then the D3-brane in the second line becomes an \adss{3}{1}
probe that intersects the boundary of \adss{5}{5} on a
line-defect (along the 3-direction), on which the endpoints of
the strings appear as an electric charge density.
The addition of angular momentum in the 12-plane to this probe D3/F1 system
causes it to expand to a D5-tube with geometry $\bbr{4} \times \calc$,
so the array \eqn{d3-d3-f1} is replaced by
\be
\begin{array}{rcccccccccl}
\mbox{D3:}\,\,\, & 1 & 2 & 3 & \_ & \_ & \_ & \_ & \_ & \_ & \, \\
\mbox{D5-tube:}\,\,\, & \bullet & \bullet & 3 & 4 & 5 & \underline{6} & \_ & \_ & \_ &.
\ea
\label{d3-t5}
\ee
In the second line the underlined
direction indicates the orientation of the strings dissolved within
the tube, whilst the remaining directions are those of the D3-branes
also dissolved within the tube. The bullets indicate that the
cross-section $\calc$ of the D5-tube is restricted here to be a
curve in the 12-plane (we will consider the general case in
Section~\ref{ads-stubes}). 
Since the D5-tube preserves the same supersymmetries as the
collection of strings and D3-branes from which it originates, the
intersection \eqn{d3-t5} is 1/8-supersymmetric.
If the D3-branes are now replaced by the \adss{5}{5}
background then the resulting probe D5-tube ends at the \ads{5}
boundary on a defect of geometry $\bbr{} \times \calc$ (see Figure
\ref{d3-t5-fig}). 
\FIGURE[t]{
{\epsfig{file=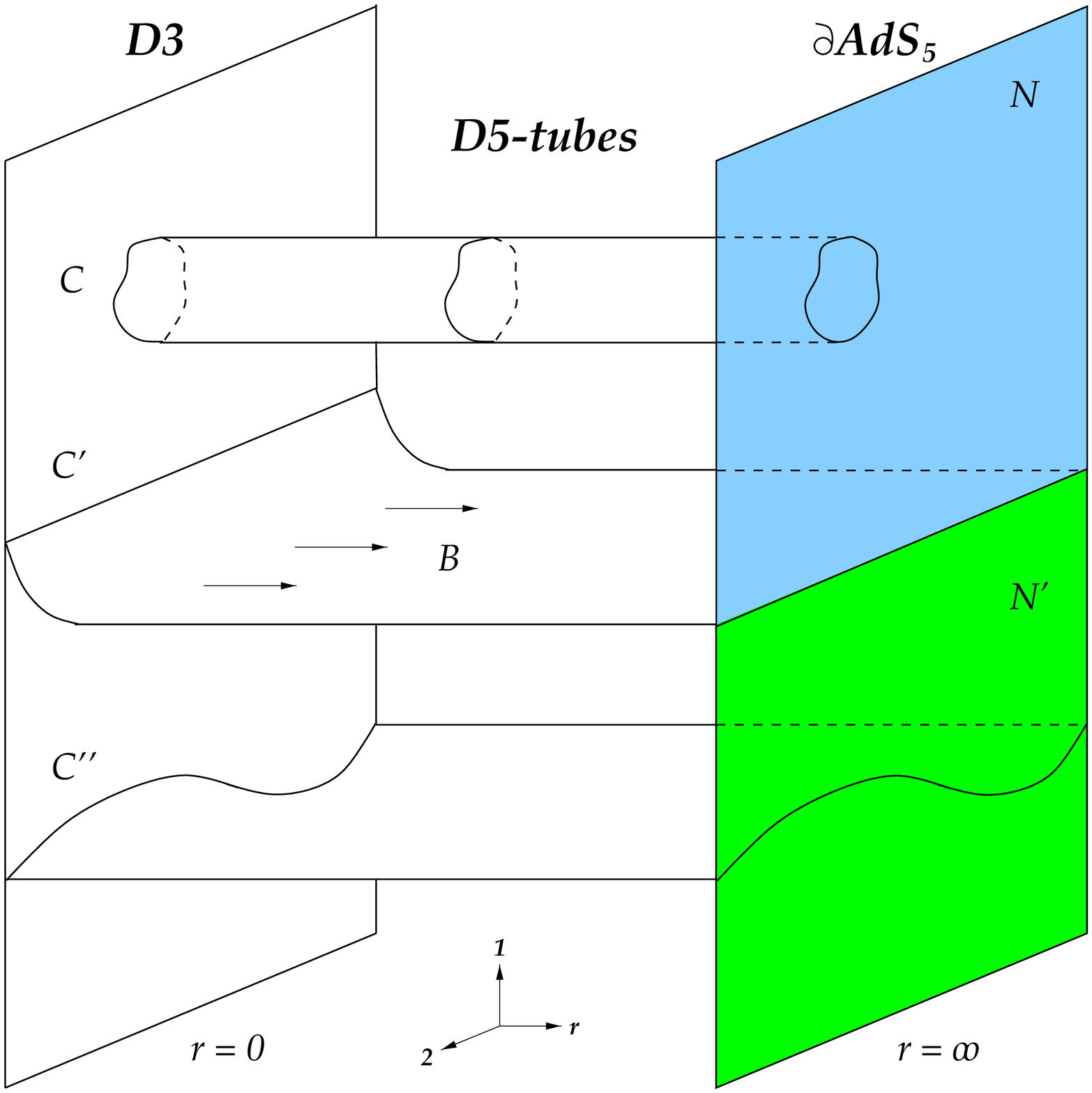, height=12cm}}
\caption{Here we have depicted three examples of D5-tubes  
intersecting D3-branes. (It should be understood that the
D5-tubes continue through to the other side of the D3-branes.) The
cross-sections of the tubes may be open ($\calc'$ and $\calc''$) or
closed ($\calc$). The numbers of D3-branes 
$N$ and $N'$ (here with $N\geq N'$) on either side of a D5-tube may
differ in the case that the 
cross-section is a straight line ($\calc'$).  In this case the
boundaries of the D3-branes deform the D5-tube and source a
magnetic BI field $B$ on it. If the D3-branes are replaced by 
an \adss{5}{5} background then the 
D5-tubes intersect the boundary of \ads{5} (at $r\ra \infty$) 
on a tubular defect with cross-section equal to that of the
corresponding tube. The gauge group of the SYM theory is 
$SU(N)$ on one side of the planar defect associated to the tube with 
linear cross-section $\calc'$, and $SU(N')$ on the other.}
\label{d3-t5-fig}
}
From the viewpoint of the \nef SYM theory we conclude 
that the addition of electric charge and angular momentum to a 
line-defect causes it to expand to a defect of one dimension higher 
with geometry $\bbr{} \times \calc$ while leaving unbroken four of the eight
Poincar\'e supersymmetries preserved by the flat defect; conformal
(super)symmetry is of course completely broken for non-linear $\calc$.
Analogous arguments show that a similar expansion occurs for the
defects in columns I and II of Table \ref{defects}, with the exception of those involving D7-branes.

\TABLE[hhh]{
\vspace{.2cm}
\begin{tabular}{|c|c|}
\hline 
I & II \\
\hline 
\dd{0}{3}{1} & \dd{0}{3}{5} \\ 
\dd{1}{3}{3} & \dd{1}{3}{7} \\
\dd{2}{3}{5} & \\
\dd{3}{3}{7} & \\
\hline
\end{tabular}
\caption{{\it 1/4-supersymmetric} D3/D$p$ {\it intersections.}}
\label{defects}
}

An \bbr{q}-defect can thus be regarded as an expanded \bbr{q-1}-defect
in the limit in which $\calc$ becomes a straight line; if in addition
the electric and magnetic BI fields on the dual D-tube are set to zero
then conformal (super)symmetry is restored and the resulting
\bbr{q}-defect reduces to one of the defects in Table \ref{defects}.
Similarly, an \bbr{q-1}-defect may be 
viewed as a collapsed $q$-dimensional defect in the limit in 
which $\calc$ reduces to a point and the BI fields are set to zero. 
Thus all defects associated to the D3/D$p$ intersections in column I 
may be considered in a unified way\footnote{The last entry in column I
is not really associated to 
a defect because the intersection has codimension zero inside the
D3-brane; rather, adding a probe D7-brane is associated to introducing
dynamical `quarks' (or, more precisely, matter in the fundamental
representation of the gauge group) in the gauge theory \cite{KK02}.};
the same applies to those defects in column II.   We will see in
Section~\ref{discussion} how defects in columns I and II are related
by considering configurations with multiple defects.   

The \bbr{2}-defect is special because, being of codimension one, it
can separate two {\it different} CFTs defined on either side of it;
in particular, the ranks of their gauge groups may differ. In Section
\ref{gauge-jumps} we investigate whether this is also possible for the
non-planar $(\bbr{} \times \calc)$-defects uncovered here.

%%%%%%%%%%%%%%%%%%%%%%%%%%%%%%%%%%%%%%%%%%%%%%%%%%%%%%%%%%%%%%%%%%%%%%%%%%%%%%
%%%%%%%%%%%%%%%%%%%%%%%%%%%%%%%%%%%%%%%%%%%%%%%%%%%%%%%%%%%%%%%%%%%%%%%%%%%%%%
%%%%%%%%%%%%%%%%%%%%%% SECTION 2 %%%%%%%%%%%%%%%%%%%%%%%%%%%%%%%%%%%%%%%%%%%%%
%%%%%%%%%%%%%%%%%%%%%%%%%%%%%%%%%%%%%%%%%%%%%%%%%%%%%%%%%%%%%%%%%%%%%%%%%%%%%%
%%%%%%%%%%%%%%%%%%%%%%%%%%%%%%%%%%%%%%%%%%%%%%%%%%%%%%%%%%%%%%%%%%%%%%%%%%%%%%
%\newpage
\sect{Anti-de Sitter Supertubes}
\label{ads-stubes}
%%%%%%%%%%%%%%%%%%%%%%%%%%%%%%%%%%%%%%%%%%%%%%%%%%%%%%%%%%%%%%%%%%%%%%%%%%%%%%
%%%%%%%%%%%%%%%%%%%%%%%%%%%%%%%%%%%%%%%%%%%%%%%%%%%%%%%%%%%%%%%%%%%%%%%%%%%%%%

In this section we will explicitly verify that a D5-tube embedded in a
D3-brane background with the orientation
\be
\begin{array}{rcccccccccl}
\mbox{D3}:\,\,\, & 1 & 2 & 3 & \_ & \_ & \_ & \_ & \_ & \_ & \, \\
\mbox{D5-tube:}\,\,\, & \bullet & \bullet & 3 & 4 & 5 & \underline{6} & 
\bullet & \bullet & \bullet &
\ea
\label{d3-t5-gen}
\ee
indeed preserves a fraction of supersymmetry, as stated in the
Introduction; the calculation for any other D$p$-tube is completely
analogous. Note that \ref{d3-t5-gen} generalizes \eqn{d3-t5} in that the
cross-section of the D5-tube is now a completely arbitrary curve in
the \bbr{5}-space corresponding to the 12789-directions; this is the
D5-tube to which the D3/F1 system in \eqn{d3-d3-f1} expands for a 
general angular momentum two-form in this \bbr{5}-space.

The non-zero fields of the supergravity solution sourced by D3-branes 
take the form
\bea
ds^2 &=& H^{-1/2} ds^2 \left( \bbe{(1,3)} \right) + 
H^{1/2} ds^2 \left( \bbe{6} \right) \,, \nn
F_\ifive &=& \omega + * \omega \sac 
\omega = vol \left(\bbe{(1,3)} \right) \w dH^{-1} \sac 
e^\phi = g_s \,,
\label{d3}
\eea
where $H$ is a harmonic function on $\bbe{6}$. If $H=1$ this background
reduces to flat space; if instead $H=1+ R^4 / r^4$, 
where $R^4=g_s N$ and $r$ is the radial coordinate in $\bbe{6}$, 
then it describes $N$ D3-branes at $r=0$. 
The near-horizon limit is effectively obtained
by removing the `1' in the harmonic function, in which case the geometry
becomes \adss{5}{5}. For the moment we need not choose a specific form
for $H$ because most of the results we will obtain do not depend on it.

In view of \eqn{d3-t5-gen} we denote the coordinates in
$\bbe{(1,3)}$ as $\bx=(x^0,x^3)$ and $\bX=(X^a)$, where $a=1,2$, and
for those in $\bbe{6}$ we use $\by=(y^m)$, where $m=4,5,6$, 
and $\bY=(Y^7,Y^8,Y^9)$. Since the D5-tube extends along the 
03456-directions, we identify $\bx$ and $\by$ with worldvolume
coordinates. The additional worldvolume coordinate $\s$ parametrizes 
the cross-section $\calc$, which is specified by
\be
\bX = \bX(\s) \sac \bY= \bY(\s) \,.
\label{xy}
\ee
The BI field strength on the D5-tube takes the form
\be
F = \cale\, dx^0 \w dy^6 + \calb \, dy^6 \w d\s \,.
\label{F}
\ee
The electric and magnetic components $\cale$ and $\calb$ correspond,
respectively, to fundamental strings along the 6-direction and
D3-branes along the 345-directions dissolved in the D5-brane. The
Poynting momentum-density generated by the crossed electric and
magnetic fields is responsible for the stability of the tube with an
arbitrary cross-section \cite{MT01,MNT01}.

The supersymmetries preserved by a D5-brane are 
those generated by Killing spinors $\e$ of the background 
that satisfy (see, for example, \cite{BKOP97})
\be
\G \e = \e \,,
\label{susy}
\ee
where $\G$ is the matrix appearing in the kappa-symmetry
transformations of the D5-brane worldvolume fermions \cite{CGNSW96,BT96}. For the 
case of interest here we have 
\be
\Delta\, \G = \calb \, \G_{0345}I - \g \G_{345} \left(\G_{06}K - \cale \right)
\,, 
\ee
where $\{\G_0, \ldots ,\G_9\}$ are ten flat-space constant Dirac
matrices, $K$ and $I$ are operators that act on $SO(1,9)$ chiral 
complex spinors as 
\be
K \psi = \psi^* \sac I \psi = -i \psi \,,
\ee
and $\Delta=\sqrt{-\det(g+F)}$ is the DBI determinant, which in the 
present case is given by
\be
\Delta=\sqrt{\calb^2 + (1-\cale^2)(|\pa_\s \bX|^2 + H|\pa_\s \bY|^2)}\,.
\ee
We have also introduced 
\be
\g = \sum_{i=1}^2 \G_i \pa_\s X^i + H^{1/2} \sum_{i=7}^9 \G_i \pa_\s Y^i \,.
\ee
For generic $H$ the Killing spinors of the background \eqn{d3} take the
form  
\be
\e = H^{-1/8} \e_0 \,,
\label{killing}
\ee
where $\e_0$ is a constant spinor subject to the constraint 
\be
\G_{0123} I \e_0 = \e_0 
\label{d3-constraint}
\ee
associated to the presence of D3-branes along the 123-directions.
This means that the background is
invariant under sixteen real supersymmetries. This is enhanced to
thirty-two supersymmetries in two particular cases. If $H=1$ then the
background is just the Minkowski vacuum and $\e_0$ is unconstrained. If
$H=R^4/r^4$ then the background is \adss{5}{5} and there are sixteen
special conformal supersymmetries in addition to the Poincar\'e
supersymmetries generated by $\e_0$. We will not need the specific form of the
conformal supersymmetries in this section. 

Equation \eqn{susy} is satisfied for {\it arbitrary} cross-section if 
\be
|\cale| =  1 \,,
\label{bps-e}
\ee
$\calb$ is a constant-sign function on the D5-brane worldvolume, 
and $\e_0$ satisfies the additional constraints 
\be
\G_{06} K \e_0 = \sgn{\cale} \e_0 \sac \G_{0345} I \e_0 = \sgn{\calb} \e_0 \,.
\label{proj}
\ee
These are respectively associated to the presence of string 
charge in the 6-direction and \mbox{D3-brane} charge in the 345-directions, 
as expected from the charges carried by the D5-tube. Since these
constraints are compatible with each other and with \eqn{d3}, the D5-tube
preserves four of the sixteen supersymmetries of the D3-brane 
background\footnote{
If $H=1$ the D5-tube preserves sixteen of the thirty-two Minkowski
supercharges; the case of \adss{5}{5} is discussed below.}. 
Note that, as anticipated, there is no projection 
corresponding to the presence of the D5-brane.
Although supersymmetry allows $\calb$ to depend on all the
worldvolume coordinates, the Bianchi identity $dF=0$ restricts it to
depend only on $y^6$ and $\s$, and the equations of motion for $F$
further disallow the dependence on $y^6$. Hence we conclude that $\calb$
may be a constant-sign but otherwise arbitrary function of
$\s$.\footnote{
We have verified that all the remaining D5-brane equations of motion
are satisfied under these circumstances.}  
We would like to emphasize that if $\calb \neq 0$ then $|\cale|=1$ is a
{\it sub}critical electric field, by which we mean that the DBI
Lagrangian is real and nonzero.  
As anticipated, the BPS equations (that is, the conditions on 
$\cale$ and $\calb$) and the projections \eqn{proj} we 
have derived for preservation of supersymmetry are independent of the
form of $H$. 

The supersymmetry preserved by the D5-tube is enhanced if the 
cross-section is a straight line in the 12-plane and $\calb$ is
constant: in this case the D5-tube preserves eight
supersymmetries for any (subcritical) constant value of $\cale$,
because the matrix $\Gamma$ becomes a constant matrix that
commutes with that in equation \eqn{d3-constraint}.

%%%%%%%%%%%%%%%%%%%%%%%%%%%%%%%%%%%%%%%%%%%%%%%%%%%%%%%%%%%%%%%%%%%%%%%%%%%%%%
%%%%%%%%%%%%%%%%%%%%%%%%%%%%%%%%%%%%%%%%%%%%%%%%%%%%%%%%%%%%%%%%%%%%%%%%%%%%%%
%%%%%%%%%%%%%%%%%%%%%% SECTION 3 %%%%%%%%%%%%%%%%%%%%%%%%%%%%%%%%%%%%%%%%%%%%%
%%%%%%%%%%%%%%%%%%%%%%%%%%%%%%%%%%%%%%%%%%%%%%%%%%%%%%%%%%%%%%%%%%%%%%%%%%%%%%
%%%%%%%%%%%%%%%%%%%%%%%%%%%%%%%%%%%%%%%%%%%%%%%%%%%%%%%%%%%%%%%%%%%%%%%%%%%%%%
%\newpage
\sect{Features of Expanded Defects}
\label{expansion}
%%%%%%%%%%%%%%%%%%%%%%%%%%%%%%%%%%%%%%%%%%%%%%%%%%%%%%%%%%%%%%%%%%%%%%%%%%%%%%
%%%%%%%%%%%%%%%%%%%%%%%%%%%%%%%%%%%%%%%%%%%%%%%%%%%%%%%%%%%%%%%%%%%%%%%%%%%%%%

We will now analyze in more detail the features of the expanded
defects in the \nef SYM theory uncovered here, so we set 
$H=R^4/r^4$. Again we concentrate on the D5-tube for concreteness since
the extension to other D$p$-tubes is straightforward.

The probe D5-tube is embedded in \adss{5}{5} in a way 
determined by the cross-section $\calc$. The two particular cases in which
$\calc$ lies entirely along the 12-directions, which we
refer to as `expansion along \ads{5}', or along the 789-directions,
which we refer to as `expansion along $S^5$', are especially
interesting.

%%%%%%%%%%%%%%%%%%%%%%%%%%%%%%%%%%%%%%%%%%%%%%%%%%%%%%%%%%%%%%%%%%%%%%%%%%%%%%
\subsection{Expansion along \ads{5}}
%%%%%%%%%%%%%%%%%%%%%%%%%%%%%%%%%%%%%%%%%%%%%%%%%%%%%%%%%%%%%%%%%%%%%%%%%%%%%%

In this case $\bY$ is constant and the D5-tube intersects the 
boundary of \ads{5} on a defect of geometry $\bbr{} \times \calc$, 
where $\calc$ is a curve in the 12-plane. The induced metric on the
D5-tube is 
\be
ds^2 = \fc{y^2 + Y^2}{R^2} \left( -dx_0^2 + d\s^2 + dx_3^2 \right)
+ \fc{R^2}{y^2 + Y^2} dy^2 + 
\fc{R^2 y^2}{y^2 + Y^2}  \, ds^2 \left( S^2\right) \,, 
\label{ind1}
\ee
where $y=|\by|$, $Y=|\bY|$ and we have set $|\pa_\s \bX|^2=1$ 
without loss of generality by assuming an affine parametrization. 

Let us first suppose that $Y=0$. Although in this case the metric is
that of \adss{4}{2}, this does {\it not} imply that part of the conformal
isometries of \ads{5} are preserved, because the submanifold
occupied by the brane is not mapped into itself by these
isometries unless $\calc$ is a straight line. From the viewpoint 
of the gauge theory it is clear that conformal symmetry 
(as well as translational and rotational symmetry in the 12-plane) 
must be broken if $\calc$ is not a straight line
since its size (or radius of curvature) introduces a scale. Similarly,
conformal supersymmetry of \adss{5}{5} is completely broken if the
cross-section of the D5-tube is not a straight line, as we show in the
Appendix. It follows that the corresponding tubular defect with
generic $\calc$ breaks all the conformal supersymmetries of the 
\nef SYM theory; of course, it preserves four of the sixteen Poincar\'e 
supersymmetries, as we saw in the previous section. 

It is thus clear that a necessary condition for preservation of
conformal symmetry is that the cross-section of the D5-tube be a
straight line, in which case the submanifold occupied by the D5-tube is
invariant under the conformal isometries of \ads{5}. However, this
condition is not sufficient because the BI field strength on the 
worldvolume of the D5-tube is not invariant\footnote{
Technically this means that its Lie derivative along the Killing 
vector fields that generate these isometries is non-zero.}
unless $\cale$ and $\calb$ are set to zero (recall that precisely for a
linear cross-section in the 12-plane this can be done while preserving
Poincar\'e supersymmetry). We thus conclude that (part of) the
conformal symmetries of \ads{5} are preserved by the D5-tube if and  
only if its cross-section is a straight line and $\cale=\calb=0$; we show
in the Appendix that the same applies for preservation of conformal
supersymmetry. Under these circumstances the D5-tube reduces to 
an ordinary D5-brane, the intersection \eqn{d3-t5-gen} becomes a 
\dd{2}{3}{5} intersection, and the corresponding \bbr{2}-defect 
preserves (part of) the conformal invariance of the \nef SYM theory.

If $Y \neq 0$ then the metric \eqn{ind1} is that of \adss{4}{2} 
for $y \gg Y$ but deviates from it in the interior of \ads{5}. 
This means that the theory on the defect is not conformal even if
$\calc$ is a straight line and $\cale=\calb=0$. In this case conformal
invariance is broken because for non-zero $Y$ the D3-D5 open strings, 
and hence the corresponding degrees of freedom on the defect, 
are massive\footnote{
Note that this breaking of conformal invariance is explicit and
supersymmetry-preserving, as opposed to the {\it
non-}supersymmetry-preserving breaking by vacuum expectation 
values discussed in \cite{ST02}.}.
The fact that the embedding becomes conformal for 
large $y$ reflects the fact that these masses become negligible in 
the ultra-violet limit.

%%%%%%%%%%%%%%%%%%%%%%%%%%%%%%%%%%%%%%%%%%%%%%%%%%%%%%%%%%%%%%%%%%%%%%%%%%%%%%
\subsection{Expansion along $S^5$}
%%%%%%%%%%%%%%%%%%%%%%%%%%%%%%%%%%%%%%%%%%%%%%%%%%%%%%%%%%%%%%%%%%%%%%%%%%%%%%

Here we shall directly concentrate on the case in which $\calc$ is a
straight line. Since $\bX$ is constant the induced metric on the
D5-tube is
\be
ds^2 = \fc{r^2}{R^2} \left( -dx_0^2 + dx_3^2 \right) +
\fc{R^2}{r^2} dr^2 + R^2 \, ds^2 \left( S^3\right) \,. 
\label{geom}
\ee
Although the submanifold occupied by the D5-tube is indeed invariant 
under part of the conformal symmetries of \ads{}, as suggested by 
the \adss{3}{3} form of the induced metric, these symmetries are again
broken by the BI field strength. Note that setting the BI fields to
zero is now incompatible with supersymmetry, since a pure D5-brane
embedded as an \adss{3}{3} submanifold is not supersymmetric\footnote{
Technically this manifests itself in the fact that the matrix 
$\Gamma$ anti-commutes with that in equation \eqn{d3} if the
cross-section lies in the 789-space and $\cale=\calb=0$.} \cite{ST02},
as could be expected from the fact that it would seem to originate
from a \dd{1}{3}{5} intersection. It is precisely the non-zero 
worldvolume electromagnetic fields which make supersymmetry 
possible in this case. The dual defect is a supersymmetric
non-conformal line-defect.

%%%%%%%%%%%%%%%%%%%%%%%%%%%%%%%%%%%%%%%%%%%%%%%%%%%%%%%%%%%%%%%%%%%%%%%%%%%%%%
%%%%%%%%%%%%%%%%%%%%%%%%%%%%%%%%%%%%%%%%%%%%%%%%%%%%%%%%%%%%%%%%%%%%%%%%%%%%%%
%%%%%%%%%%%%%%%%%%%%%% SECTION 4 %%%%%%%%%%%%%%%%%%%%%%%%%%%%%%%%%%%%%%%%%%%%%
%%%%%%%%%%%%%%%%%%%%%%%%%%%%%%%%%%%%%%%%%%%%%%%%%%%%%%%%%%%%%%%%%%%%%%%%%%%%%%
%%%%%%%%%%%%%%%%%%%%%%%%%%%%%%%%%%%%%%%%%%%%%%%%%%%%%%%%%%%%%%%%%%%%%%%%%%%%%%
%\newpage
\sect{D3-branes Ending on D5-tubes}
\label{gauge-jumps}
%%%%%%%%%%%%%%%%%%%%%%%%%%%%%%%%%%%%%%%%%%%%%%%%%%%%%%%%%%%%%%%%%%%%%%%%%%%%%%
%%%%%%%%%%%%%%%%%%%%%%%%%%%%%%%%%%%%%%%%%%%%%%%%%%%%%%%%%%%%%%%%%%%%%%%%%%%%%%

The \bbr{2}-defect dual to the \dd{2}{3}{5} intersection permits the
CFTs on either side of it to be different; the case of interest here
is when the ranks of the gauge
groups, $N$ and $N'$, differ. In the brane construction, this means 
that some D3-branes actually end on the D5-brane rather than 
intersect it, so that the difference in the numbers of D3-branes 
on either side is $\d N\equiv N-N'\neq 0$. This imbalance has
two effects on the D5-brane probe \cite{KR01} (see Figure~\ref{d3-t5-fig}). 
Firstly, it induces $\d N$ units of BI magnetic flux
through the two-sphere in the 456-space,
because the two-dimensional boundary of a D3-brane acts as a
magnetic monopole charge in the five-dimensional worldspace of the
D5-brane. Secondly, the D5-brane worldspace is deformed close to the 
boundaries of the D3-branes due to the imbalance in the forces exerted
by their tensions.

Given the existence of supersymmetric (that is, stable) tubular 
defects with arbitrary cross-section, the question of whether the 
gauge group can also jump across these defects while preserving
supersymmetry arises. To answer this we will look for the
corresponding dual \mbox{D5-tubes} in \adss{5}{5}. We will argue that in 
the probe approximation used here these do not exist, and then discuss the 
meaning of this result and the limitations of the approximation.
 
The D5-tubes we seek should be oriented as indicated by \eqn{d3-t5}, 
so we set $\bY=\mathbf{0}$, but the cross-section specified before by $\bX(\s)$ 
should now be allowed to vary in the 456-directions to describe the 
deformation due to the 
D3-brane tension imbalance; in other words, we should now have 
$\bX=\bX(\s,\by)$.
We must also generalize the form of the BI field strength \eqn{F} to
incorporate the magnetic flux sourced by the D3-brane
boundaries. Therefore we must set  
\be
F = \cale\, dx^0 \w dy^6 + \calb \, dy^6 \w d\s +
\fc{1}{2} \e_{mnp} B^m \, dy^n \w dy^p 
\label{Fbis}
\ee
and require
\be
\d N \equiv \fc{1}{2 \pi} \int_{S^2} \vec{B} \cdot \vec{dS} \neq 0 \,,
\label{flux}
\ee
where $\vec{B} = (B^m)$ and $S^2$ is a two-sphere in the 456-space. 
Recall that $a,b, \ldots = 1,2$ and $m, n, \ldots = 4,5,6$. 
Since we will need to distinguish between the 45-directions and the 
6-direction because the symmetry between them is broken by 
$\cale$ and $\calb$, we will further set $\by=(y^m) = (y^i, y^6)$ with
$i,j =4,5$. Note that in principle we should expect $\calb$ and $B^m$
to depend on $\s$ and $y^m$, of course subject to the conditions 
\bea
\pa_m B^m &=& 0 \,, \label{bi1} \\
\pa_\s B_4 &=& \pa_5 \calb \,, \label{bi2} \\
\pa_\s B_5 &=& - \pa_4 \calb \,, \label{bi3} \\
\pa_\s B_6 &=& 0  \label{bi4}
\eea
implied by the Bianchi identity $dF=0$.

Under these circumstances the kappa-symmetry matrix takes the form
\bea
\Delta \G &=& \calb \, \G_{0345} I + 
H^{-1} B^m \, \pa_{[m} X^1 \pa_{\s]} X^2 \, \G_{0123} I \nn
&& + \left[ \G_{345} \left(\g + H^{-1/2} \pa_{[i} X^1 \pa_{\s]} X^2 \G_{i12}
\right) I  + H^{-1/2} B_6 \G_3 \, \g \, KI \right]
\left(\G_{06} K - \cale \right) \nn
&& + \G_{03} KI \left( 
\calb B_6 + \G_{1245}\, \pa_{[6} X^1 \pa_{\s]} X^2 \right) 
+ H^{-1/2} \left[ \calb\, \pa_i X^a \G_{ai} \G_{0345} I + 
 B_i \, \pa_\s X^{[a} \G^{b]} \G_i \G_{0123} I \right] \nn
&& + H^{-1} \left[\fc{1}{6} \G_{0123} \g KI \, \G_{[m} \pa_n X^1 \pa_{p]} X^2 +
\G_{0123}I \left( \calb - \fc{1}{3} \cale \G_0 \, \g \right) 
\pa_{[4} X^1 \pa_{5]} X^2 \right. \nn 
&& \hspace{1.5cm} + \left. \fc{1}{3}\G_{03a}KI\,  \d_{bc}
\G_{[m} \pa_n X^a  \pa_{p]} X^b \pa_\s X^c - \fc{1}{3} \cale \G_{3a}I\, 
\d_{bc} \pa_{[i} X^a \pa_{j]} X^b \pa_\s X^c \right] \,,
\eea
where
\be
\g = \pa_\s X^a \G_a \, .
\label{gamma}
\ee
It follows that equation \eqn{susy} is satisfied if we impose the
projections \eqn{d3-constraint} and \eqn{proj}, as well as the BPS
conditions \eqn{bps-e} and 
\bea
B_i \pa_\s X^a &=& -\calb \, \e^{ab} \pa_i X^b \,, \label{bps1} \\
\calb\, B_6 &=& \pa_{[6} X^1 \pa_{\s]} X^2  \,, \label{bps2} \\
0 &=& \pa_{[m} X^1  \pa_{n]} X^2  \,, \label{bps3} 
\eea
in which case the DBI determinant reduces to
\be
\Delta = \left|\calb + H^{-1} B^m \, \pa_{[m} X^1 \pa_{\s]} X^2 \right| \,.
\ee
Note the non-linearity of the BPS equations, as well as the asymmetry between
the 45- and the 6-directions. 

The projections above are expected given the charges carried by the
D5-tube. In particular, we need to impose condition
\eqn{d3-constraint} even if $H=1$, that is, even if there are no 
background D3-branes, because these should arise here as a `spike' 
excitation on the D5-tube described by $\bX(\s,\by)$.

It is well-known that the rank of the gauge group may change while
preserving supersymmetry across an \bbr{2}-defect on which 
$\cale=\calb=0$ \cite{KR01}. Let us therefore first show that this is 
also posssible with non-zero BI fields. By rotational invariance in
the 12-plane, we assume without loss of generality that the linear 
cross-section is aligned with the 2-axis, that is, we set
\be
X^2 \equiv \s \sac X^1 \equiv X(\by) \,.
\ee
The BPS equations then simplify to
\be
B_i = \calb\,\pa_i X \sac \calb \, B_6 = \pa_6 X \,.
\label{simplify}
\ee
Under these circumstances the equations of motion for the BI gauge field 
on the D5-brane in the background \eqn{d3} reduce to the conditions
that $\calb$ is constant and that 
\be
\pa_4^2 X + \pa_5^2 X + \calb^{-2} \pa_6^2 X = 0 \, .
\ee
The latter equation coincides with the divergence-free condition
\eqn{bi1} for $B^m$ upon using \eqn{simplify}; since $\calb$ is
constant, it is solved by any
harmonic function of $\tilde{\by}$, where $\tilde{y}^i = y^i$ and
$\tilde{y}^6=\calb y^6$. The choice \mbox{$X=\d N /4 \pi |\tilde{\by}|$}
satisfies \eqn{flux}, and represents $\d N$ open D3-branes
ending on the D5-tube at $\by=\mathbf{0}$. 

We now turn to non-linear cross-sections. We will show that the
cross-section cannot be non-linear unless $\d N=0$.  It will suffice
to prove this for a circular cross-section since any curve is
approximated locally by a circle, and the condition for preservation of
supersymmetry \eqn{susy} is local.   Let us therefore set  
\be
X^1 = R(\by) \sin\s \sac X^2 = R(\by) \cos\s \, .
\ee
The BPS equations \eqn{bps1}--\eqn{bps3} then reduce to
\be
R \, B_i = \calb \, \pa_i R \sac \calb \, B_6 = R \, \pa_6 R \,.
\ee
Since $R$ is $\s$-independent, these equations together with
\eqn{bi2}--\eqn{bi4} imply that $B^m$ cannot depend on $\s$, and also that $\calb$
may only depend on $y^6$. The D5-brane equations of motion then force
$R$ to take the form
\be
R(\by) = f(y^4,y^5) \, |\calb(y^6)|^{1/2} \,,
\ee
where $f$ and $\calb$ must satisfy
\be
\left(\pa_4^2 + \pa_5^2 \right) \log f = \lambda \, f^2 \sac 
\pa_6^2 \log \calb = - 2\lambda \, \calb \,,
\label{fg}
\ee
for some undetermined constant $\lambda$. We seek a solution for which
$R$ tends to a non-zero constant far away from the D3-brane
boundaries, which is at $y\ra\infty$ if we assume 
(without loss of generality) that the
D3-branes end on the D5-tube at $y=0$. Since this means that the
left-hand sides of \eqn{fg} vanish asymptotically, we must set
$\lambda=0$. If we additionally restrict ourselves to solutions invariant
under rotations in the 45-plane, then 
\be
f = a \left(y_4^2 + y_5^2\right)^b \sac \calb = c \, e^{d y^6} \,,
\ee
for constants $a,b,c,d$. The requirement that $R$ tends to a 
{\it non-zero} constant implies that $c=d=0$, in which case $R$ is
everywhere constant and $B^m=0$. This is therefore a particular case
of the general D5-tube with an arbitrary but $\by$-independent
cross-section and $\d N=0$. 

Although the above result was derived for a particular ansatz for the
BI fields \eqn{Fbis}, this ansatz is grounded in a physical
understanding of supertubes, and hence we consider it unlikely that
there exists a supersymmetric solution not captured by this ansatz.
Another caveat is that effects beyond the probe approximation could
allow a supersymmetric non-linear cross-section provided that its radius of
curvature is of the order $N/\d N$, consistent with the fact that in
the limit $\d N/N\ra 0$ only linear cross-sections are supersymmetric.

%%%%%%%%%%%%%%%%%%%%%%%%%%%%%%%%%%%%%%%%%%%%%%%%%%%%%%%%%%%%%%%%%%%%%%%%%%%%%%
%%%%%%%%%%%%%%%%%%%%%%%%%%%%%%%%%%%%%%%%%%%%%%%%%%%%%%%%%%%%%%%%%%%%%%%%%%%%%%
%%%%%%%%%%%%%%%%%%%%%% SECTION 5 %%%%%%%%%%%%%%%%%%%%%%%%%%%%%%%%%%%%%%%%%%%%%
%%%%%%%%%%%%%%%%%%%%%%%%%%%%%%%%%%%%%%%%%%%%%%%%%%%%%%%%%%%%%%%%%%%%%%%%%%%%%%
%%%%%%%%%%%%%%%%%%%%%%%%%%%%%%%%%%%%%%%%%%%%%%%%%%%%%%%%%%%%%%%%%%%%%%%%%%%%%%
%\newpage
\sect{Multiple Defects}
\label{discussion}
%%%%%%%%%%%%%%%%%%%%%%%%%%%%%%%%%%%%%%%%%%%%%%%%%%%%%%%%%%%%%%%%%%%%%%%%%%%%%%
%%%%%%%%%%%%%%%%%%%%%%%%%%%%%%%%%%%%%%%%%%%%%%%%%%%%%%%%%%%%%%%%%%%%%%%%%%%%%%

In this section we discuss configurations with multiple defects, some of
which involve defects from both columns I and II in Table \ref{defects}. 

A first observation is that any number of defects with geometries
$\bbr{q} \times \calc_i$ (for fixed $q$), where the \bbr{q}-factors
are mutually-aligned and the cross-sections $\calc_i$ are arbitrary,  
preserve 1/4 of the Poincar\'e supersymmetries of the \nef SYM theory.
(Three of them with $q=1$ have been depicted in Figure \ref{d3-t5-fig}.) 
This follows immediately from the fact that D$p$-tubes with
mutually-aligned axes and arbitrary cross-sections preserve
1/4-supersymmetry, as was shown in \cite{MNT01} by exhibiting a
1/4-supersymmetric supergravity solution representing such a
multi-tube configuration\footnote{
For circular cross-sections the supergravity solution was first 
found in \cite{EMT01}, and solutions in Matrix theory were constructed
in \cite{BL01}.}. 
Note that this includes as a particular case parallel defect/anti-defect
configurations, for which the dual system consists of a  supersymmetric parallel
brane/anti-brane pair \cite{MNT01, BK01}.

We have shown that the line-defect of column I can be viewed as a
collapsed tubular two-dimensional defect. This suggests that the
line-defect could have an endpoint on a planar \bbr{2}-defect, or
that it could even be smoothly attached to it. The
dual configuration in \adss{5}{5} that realizes this possibility is
described by the array
\be
\begin{array}{rcccccccccl}
\mbox{D3:}\,\,\, & 1 & 2 & 3 & \_ & \_ & \_ & \_ & \_ & \_ & \, \\
\mbox{D5:}\,\,\, & \_ & 2 & 3 & 4 & 5 & 6 & \_ & \_ & \_ & \, \\
\mbox{D3:}\,\,\, & 1 & \_ & \_ & 4 & 5 & \_ & \_ & \_ & \_ & \, .
\ea
\label{d3/d5/d3}
\ee
As usual the D3-branes at the top provide the \adss{5}{5} background. 
The probe D3-brane at the bottom ends on the probe D5-brane, that
is, it extends only along the (say) positive values of the
1-coordinate. From the viewpoint
of the SYM theory this represents a  line-defect ending
on an \bbr{2}-defect. Since the D3-brane can be viewed as a smooth solitonic
excitation of the D5-brane \cite{CM97,Gibbons97}, one might expect
that the line defect can 
be similarly viewed as a solitonic excitation of the conformal field
theory on the \bbr{2}-defect as depicted in Figure~\ref{line-on-plane}.

\FIGURE{
{\epsfig{file=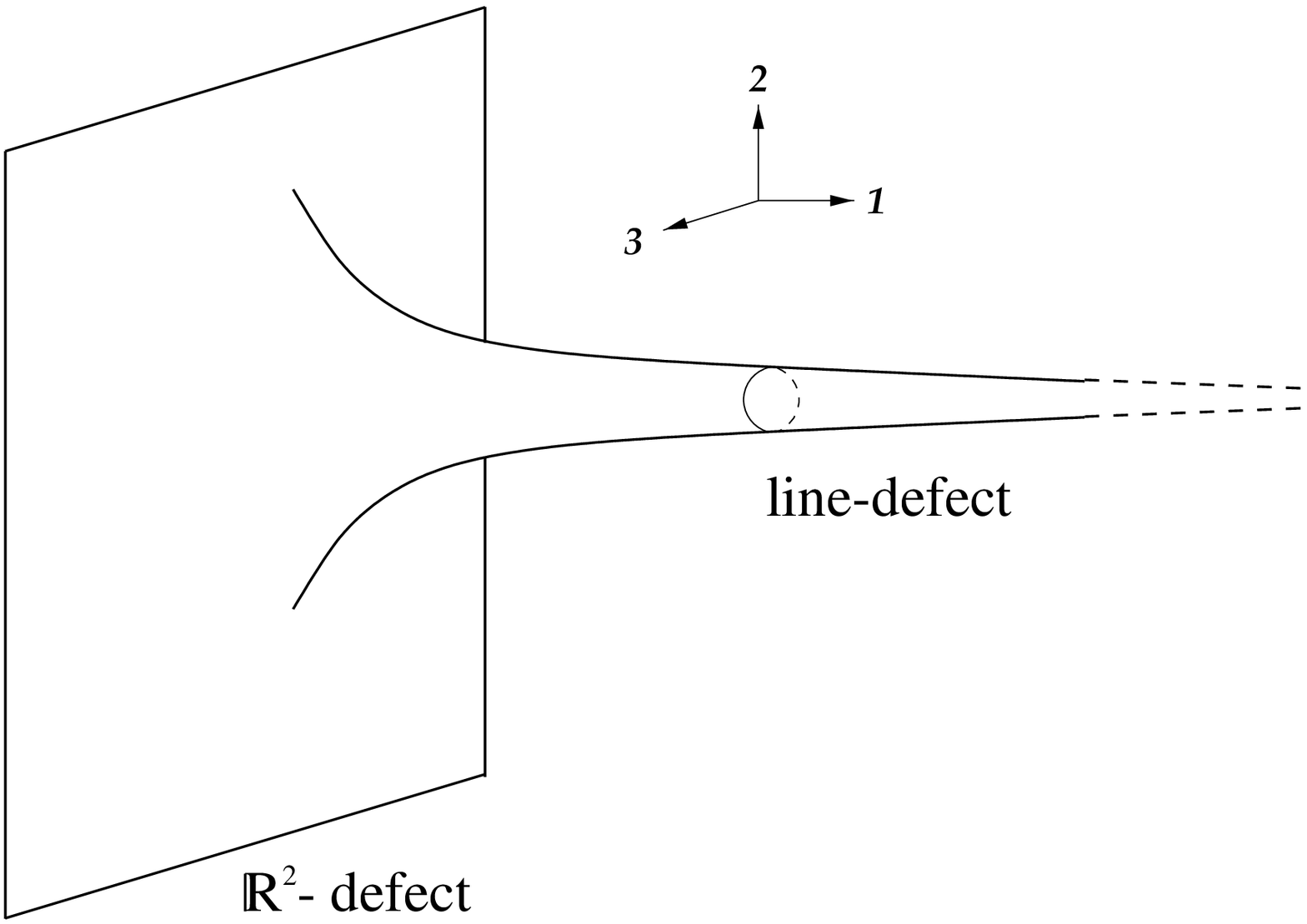, height=7cm}}
\caption{A line-defect ending on an \bbr{2}-defect.}
\label{line-on-plane}
}

The CFT on the \bbr{2}-defect associated to the \dd{2}{3}{5}
intersection was deduced in \cite{DFO01}, following the earlier work of
\cite{Sethi97,KS98}. 
Essentially, it is a linear \nef $d=3$ sigma-model with $4N$ scalar
fields but with additional couplings to the pullbacks of the $d=4$
SYM fields\footnote{The same field theory, but in a lower dimension, should
therefore give the dynamics of the line- and point-defects associated to the 
\dd{1}{3}{3} and \dd{0}{3}{1} intersections because each of them  
can be regarded as a collapsed \bbr{2}-defect.}. A vortex solution of
this theory could interpolate between a region of non-zero scalar
fields at the vortex core
to an asymptotic region of vanishing scalar fields far away from the core.
Due to the coupling to the four-dimensional background fields,
non-vanishing scalar fields on the defect imply discontinuities of the
Higgs fields across the defect.  To take this properly into account
would require going beyond the probe approximation. 

\FIGURE[t]{
{\epsfig{file=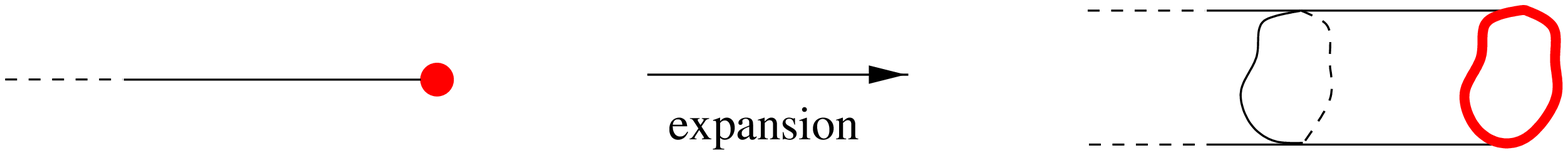, height=1.5cm}}
\caption{On the left-hand side a line-defect ends on a point-defect;
on the right-hand side the expanded line-defect ends
on the expanded point-defect after the addition of angular momentum.}
\label{tube-tube}
}

Given that a line-defect can end on a planar defect, one may wonder
whether a tubular defect formed by expansion of a line-defect can also
end on a planar defect. 
This would seem
to require a corresponding bulk configuration in which a D5-tube has a
boundary on a D5-brane, and there is no known supersymmetric
configuration of this kind; a possible resolution
is that the D5-brane expands {\it locally} to a D7-tube, on which the D5-tube
can end. 
This illustrates a rather general problem in any attempt to find a
supertube {\it ending} on another brane, as opposed to merely intersecting it. 
An examination of all possibilities \cite{MN01}
shows that the only case in which the problem does not arise is that recently 
discussed \cite{KMPW02} of a D2-tube ending on a D4-brane, and
the T-dual cases in which a D$p$-tube ends on a D$(p+2)$-brane. Note
that a naive application of T-duality rules to systems with branes
ending on branes can lead to incorrect results. For example, 
T-duality along a direction longitudinal to the D4-brane and
orthogonal to the D2-brane that ends on it might seem to transform this
system into one in which a D3-brane has a one-dimensional boundary (on
another D3-brane), which is impossible.

The line-defect of column I in Table \ref{defects} can also end on the
point-defect associated to the \dd{0}{3}{5} intersection of column
II (see Figure \ref{tube-tube}). The corresponding brane configuration is
\be
\begin{array}{rcccccccccl}
\mbox{D3:}\,\,\, & 1 & 2 & 3 & \_ & \_ & \_ & \_ & \_ & \_ & \, \\
\mbox{D5:}\,\,\, &\_ &\_ &\_ & 4 & 5 & 6 & 7 & 8 & \_ & \, \\
\mbox{D3:}\,\,\, & \_ & \_ & 3 & 4 & 5 & \_ & \_ & \_ & \_ & \,,
\ea
\label{d3/d5'/d3}
\ee
where the probe D3-brane at the bottom ends on the probe D5-brane.
If strings along the 9-direction and angular momentum in the 12-plane
are added, then both probes can expand supersymmetrically yielding a configuration in
which a D5-tube ends on a D7-tube: 
\be
\begin{array}{rcccccccccl}
\mbox{D3:}\,\,\, & 1 & 2 & 3 & \_ & \_ & \_ & \_ & \_ & \_ & \, \\
\mbox{D7-tube}:\,\,\, &
\bullet &\bullet & \_ & 4 & 5 & 6 & 7 & 8 & \underline{9} & \, \\
\mbox{D5-tube}:\,\,\, & 
\bullet & \bullet & 3 & 4 & 5 & \_ & \_ & \_ & \underline{9} & \,,
\ea
\ee
Both tubes share a common cross-section $\calc$ in the 12-plane. From
the viewpoint of the CFT theory, a tubular 
$\bbr{} \times \calc$-defect with axis along the 3-direction ends on a
one-dimensional $\calc$-defect (see Figure \ref{tube-tube}). In the
limit in which $\calc$ becomes a straight line, the $\calc$-defect
reduces to the line-defect associated to the second intersection in
column II of Table \ref{defects}, and the configuration above becomes
an \bbr{2}-defect with a boundary line-defect. For a line in the (say)
2-direction this corresponds to the array
\be
\begin{array}{rcccccccccl}
\mbox{D3:}\,\,\, & 1 & 2 & 3 & \_ & \_ & \_ & \_ & \_ & \_ & \, \\
\mbox{D7:}\,\,\, &\_ & 2 &\_ & 4 & 5 & 6 & 7 & 8 & 9 & \, \\
\mbox{D5:}\,\,\, &\_ & 2 & 3 & 4 & 5 & \_ & \_ & \_ & 9 & \,.
\ea
\ee

%%%%%%%%%%%%%%%%%%%%%%%%%%%%%%%%%%%%%%%%%%%%%%%%%%%%%%%%%%%%%%%%%%%%%%%%%%%%%%
%%%%%%%%%%%%%%%%%%%%%%%%%%%%%%%%%%%%%%%%%%%%%%%%%%%%%%%%%%%%%%%%%%%%%%%%%%%%%%
%%%%%%%%%%%%%%%%%%%%%% ACKNOWLEDGMENTS %%%%%%%%%%%%%%%%%%%%%%%%%%%%%%%%%%%%%%%
%%%%%%%%%%%%%%%%%%%%%%%%%%%%%%%%%%%%%%%%%%%%%%%%%%%%%%%%%%%%%%%%%%%%%%%%%%%%%%
%%%%%%%%%%%%%%%%%%%%%%%%%%%%%%%%%%%%%%%%%%%%%%%%%%%%%%%%%%%%%%%%%%%%%%%%%%%%%%
%\newpage
\acknowledgments

We are grateful to Costas Bachas and Hirosi Ooguri for helpful
conversations. D.M. is supported by a PPARC fellowship.  S.N. is
supported by the British Federation of Women Graduates and the
Australian Federation of University Women (Queensland).

%%%%%%%%%%%%%%%%%%%%%%%%%%%%%%%%%%%%%%%%%%%%%%%%%%%%%%%%%%%%%%%%%%%%%%%%%%%%%%
%%%%%%%%%%%%%%%%%%%%%%%%%%%%%%%%%%%%%%%%%%%%%%%%%%%%%%%%%%%%%%%%%%%%%%%%%%%%%%
%%%%%%%%%%%%%%%%%%%%%%%%% APPENDIX %%%%%%%%%%%%%%%%%%%%%%%%%%%%%%%%%%%%%%%%%%%
%%%%%%%%%%%%%%%%%%%%%%%%%%%%%%%%%%%%%%%%%%%%%%%%%%%%%%%%%%%%%%%%%%%%%%%%%%%%%%
%%%%%%%%%%%%%%%%%%%%%%%%%%%%%%%%%%%%%%%%%%%%%%%%%%%%%%%%%%%%%%%%%%%%%%%%%%%%%%
%\newpage
\appendix
\section{Special Conformal Supersymmetries}
\renewcommand{\theequation}{A.\arabic{equation}}
%%%%%%%%%%%%%%%%%%%%%%%%%%%%%%%%%%%%%%%%%%%%%%%%%%%%%%%%%%%%%%%%%%%%%%%%%%%%%%

If the cross-section of the D5-tube is a line in the 12-plane (at
$\bY=0$) and $\cale=\calb=0$ then eight of the sixteen special
conformal supersymmetries of \adss{5}{5} are preserved (together with
eight of the sixteen Poincar\'e supersymmetries) \cite{ST02}.  We show
here that these conditions are not only sufficient but also necessary 
for preservation of conformal supersymmetry, as
stated in Section 3.1.

It is convenient to set $R=1$ and to use spherical coordinates 
$\{r, \t_\a\}$ ($\a=1, \ldots, 5$) for the \bbr{6}-space transverse to the
D3-branes such that $\bY=0$ corresponds to $\t_1=\t_2=\t_3=\pi/2$. The
D5-tube then wraps the $S^2$ within the $S^5$ parametrized by 
$\t_4, \t_5$, and its embedding in \ads{5} is specified by $\bX(\s)$. 
In these coordinates the BI field strength \eqn{F} and the DBI
determinant take the form  
\bea
F &=& \cale\cos\t_4 \,dx_0\wedge dr - \cale r\sin\t_4 \,dx_0\wedge d\t_4 
+ \calb\cos\t_4 \,dr\wedge d\s - \calb r\sin\t_4 \,d\t_4 \wedge d\s
\,, \\
\Delta &=& r^2\sin\t_4 \sqrt{\calb^2+(1-\cale^2)|\pa_\s \bX|^2} \,,
\label{det}
\eea
whereas the kappa-symmetry matrix is
\be
\Delta \,\G = r^2\sin\t_4 \, \G_{3\t_5} I \, \Big\{ 
\left[\G_{0r\t_4}K + \cale \left( \cos\t_4\,\G_{\t_4} +
\sin\t_4\,\G_r \right) \right] \g - \calb\,\G_0 \left( \cos\t_4\,\G_{\t_4} +
\sin\t_4\,\G_r \right) \Big\} \,.
\ee
Here $\g$ is as defined in \eqn{gamma} and
$\{ \G_0, \ldots \G_3, \G_r, \G_{\t_1}, \ldots \G_{\t_5} \}$ are
ten tangent-space constant Dirac matrices. 

Following \cite{ST02} closely we write the Killing spinors for \adss{5}{5} as
\be
\e = r^{1/2} h(\t_\a) \, \eta_1 + \left[ -r^{-1/2}\G_r +
r^{1/2}( x_0\G_0 + X^a\G_a + x_3\G_3 )\right] h(\t_\a) \, \eta_2 \,,
\ee
where $\eta_1$ and $\eta_2$ are {\it constant} sixteen-component
complex spinors of negative and positive chirality respectively,
subject to the additional constraints 
\be
\G_{0123} I \, \eta_1 = \eta_1 \sac \G_{0123} I \, \eta_2 = - \eta_2 \,,
\ee
and the matrix 
\be
h(\t_\a) = \exp\left(\frac{\t_1}{2} \G_{r\t_1}\right) \,
\exp\left(\frac{\t_2}{2} \G_{\t_1\t_2}\right) \,
\exp\left(\frac{\t_3}{2} \G_{\t_2\t_3}\right) \,
\exp\left(\frac{\t_4}{2} \G_{\t_3\t_4}\right) \,
\exp\left(\frac{\t_5}{2} \G_{\t_4\t_5}\right) \,
\ee
accounts for the tangent-space rotation induced by the change from
Cartesian to spherical coordinates in \bbr{6}. If $\eta_2=0$ then the
remaining eight independent complex Killing spinors are the Poincar\'e
spinors \eqn{killing}. The Killing spinors $\e$ with $\eta_2\neq 0$
(but $\eta_1$ not necessarily zero) correspond to special conformal
supersymmetries, as indicated by their dependence on the coordinates
along the D3-branes. 

Since equation \eqn{susy} must hold at all points of the D5-tube, 
terms with different functional dependencies on the worldvolume
coordinates must vanish independently. The term with $r^{3/2}$
dependence in \eqn{susy} reduces to
\bea
& & \G_{3\t_5}I \Big\{ 
\left[-\G_{0r\t_4}K +\cale(\cos\t_4 \G_{\t_4}
-\sin\t_4\,\G_r)\right] \g -
\calb\,\G_0(\cos\t_4\,\G_{\t_4} - \sin\t_4\,\G_r)\Big\} \, h(\t_\a) \eta_2 \nn
& & \hspace{6cm} = \sqrt{\calb^2 + (1-\cale^2)|\pa_\s\bX|^2} \,
h(\t_\a) \eta_2 \, . 
\eea
The term with $x_0$ dependence similarly reduces to
\bea
& &\G_{3\t_5}I\Big\{ \left[-\G_{0r\t_4}K + \cale(\cos\t_4 \G_{\t_4}
+\sin\t_4\,\G_r)\right]\g +
\calb\,\G_0(\cos\t_4\,\G_{\t_4} + \sin\t_4\,\G_r)\Big\} \, h(\t_\a)
\eta_2 \nn
& & \hspace{6cm} = \sqrt{\calb^2+(1-\cale^2)|\pa_\s\bX|^2} \, h(\t_\a)
\eta_2 \, . 
\label{eqn-t}
\eea
Subtracting the two equations above, we see that
\be
\Big[\cale\sin\t_4 \, \G_r \g + \calb\cos\t_4 \,
\G_{0\t_4}\Big] h(\t_\a)\eta_2 = 0 \, .
\ee
Using the commutation properties of the gamma matrices to bring
$h(\t_\a)$ through to the left, we find that
\be
\left[\exp\left(\frac{\t_5}{2}\G_{\t_4\t_5}\right) \G_{\t_1} \g \,
\sin\t_4 \, \cale -
\exp\left(-\frac{\t_5}{2}\G_{\t_4\t_5}\right) \G_{0\t_4}\,
\cos\t_4 \, \calb \right] \, \eta_2 = 0 \,. 
\ee
Since the two terms must vanish independently for all $\t_4, \t_5$ and
the matrices in each term are invertible, we see that this
equation can be satisfied with non-zero $\eta_2$ only if
$\cale=\calb=0$. If we substitute this condition back into 
\eqn{eqn-t} we find that 
\be
\G_{03r\t_4\t_5}KI \, \frac{\g}{|\pa_\s \bX|} \, \tilde{\eta}= \tilde{\eta}\,,
\ee
where $\tilde{\eta} = h(\t_\a) \eta_2$. Fixing $\t_\a$ and demanding
that this equation holds for all $\s$ for some non-zero $\eta_2$ (note
that $h(\t_\a)$ is invertible) requires that the matrix 
$\g/|\pa_\s \bX|$ is constant, which in turn implies that 
$\partial_\s X^a$ are constants, that is, that the cross-section is a
straight line.

%%%%%%%%%%%%%%%%%%%%%%%%%%%%%%%%%%%%%%%%%%%%%%%%%%%%%%%%%%%%%%%%%%%%%%%%%%%%%%
%%%%%%%%%%%%%%%%%%%%%%%%%%%%%%%%%%%%%%%%%%%%%%%%%%%%%%%%%%%%%%%%%%%%%%%%%%%%%%
%%%%%%%%%%%%%%%%%%%%%% BIBLIOGRAPHY %%%%%%%%%%%%%%%%%%%%%%%%%%%%%%%%%%%%%%%%%%
%%%%%%%%%%%%%%%%%%%%%%%%%%%%%%%%%%%%%%%%%%%%%%%%%%%%%%%%%%%%%%%%%%%%%%%%%%%%%%
%%%%%%%%%%%%%%%%%%%%%%%%%%%%%%%%%%%%%%%%%%%%%%%%%%%%%%%%%%%%%%%%%%%%%%%%%%%%%%
%\newpage


\begin{thebibliography}{80}

\bibitem{KR01}
A.\ Karch and L.\ Randall, 
{\sl Open and Closed String Interpretation of SUSY CFT's on 
Branes with Boundaries}, \jhep{06}{2001}{063}, \hepth{0105132}.

\bibitem{Maldacena97}
J.\ M.\ Maldacena, {\sl The Large N Limit of Superconformal Field
Theories and Supergravity}, \atmp{2}{1998}{231}, \hepth{9711200}.

\bibitem{Sethi97}
S.\ Sethi, {\sl The Matrix Formulation of Type IIB Five-Branes},
\npb{523}{1998}{158}, \hepth{9710005}.

\bibitem{KS98}
A.\ Kapustin and S.\ Sethi, {\sl The Higgs Branch of Impurity Theories},
\atmp{2}{1998}{571}, \hepth{9804027}.

\bibitem{DFO01}
O.\ DeWolfe, D.\ Z.\ Freedman and H.\ Ooguri,
{\sl Holography and Defect Conformal Field Theories},
\prd{66}{2002}{025009}, \hepth{0111135}. 

\bibitem{ST02}
K.\ Skenderis and M.\ Taylor,
{\sl Branes in AdS and pp-wave Spacetimes}, 
\jhep{06}{2002}{025}, \hepth{0204054}.

\bibitem{MT01}
D.\ Mateos and P.\ K.\ Townsend,
{\sl Supertubes},
\prl{87}{2001}{011602}, \hepth{0103030}.

\bibitem{MNT01}
D.\ Mateos, S.\ Ng and P.\ K.\ Townsend,
{\sl Tachyons, Supertubes and Brane/Anti-Brane Systems}, 
\jhep{03}{2002}{016}, \hepth{0112054}.

\bibitem{KMPW02}
M.\ Kruczenski, R.\ C.\ Myers, A.\ W.\ Peet and D.\ J.\ Winters,
{\sl Aspects of Supertubes}, \jhep{05}{2002}{017}, \hepth{0204103}. 

\bibitem{KK02}
A.\ Karch and E.\ Katz, {\sl Adding Flavor to AdS/CFT}, 
\jhep{06}{2002}{043}, \hepth{0205236}.

\bibitem{BKOP97}
E.\ Bergshoeff, R.\ Kallosh, T.\ Ort\'\i n and G.\ Papadopoulos,
{\sl Kappa-symmetry, supersymmetry and intersecting branes},
\npb{502}{1997}{149}, \hepth{9705040}.

\bibitem{CGNSW96}
M.\ Cederwall, A.\ von Gussich, B.\ E.\ W.\ Nilsson, P.\ Sundell and
A.\ Westerberg, {\sl The Dirichlet Super-p-Branes in Ten-Dimensional
Type IIA and IIB Supergravity}, \npb{490}{1997}{179}, \hepth{9611159}.

\bibitem{BT96}
E.\ Bergshoeff and P.\ K.\ Townsend, {\sl Super D-branes}, 
\npb{490}{1997}{145}, \hepth{9611173}.

\bibitem{EMT01}
R.\ Emparan, D.\ Mateos and P.\ K.\ Townsend,
{\sl Supergravity Supertubes},
\jhep{07}{2001}{011}, \hepth{0106012}.

\bibitem{BL01}
D.\ Bak and K.\ Lee, {\sl Noncommutative Supersymmetric Tubes},
\plb{509}{2001}{168}, \hepth{0103148}.

\bibitem{BK01}
D.\ Bak and A.\ Karch,
{\sl Supersymmetric Brane-Antibrane Configurations}, 
\npb{626}{2002}{165}, \hepth{0110039}.

\bibitem{CM97}
C.\ G.\ Callan and J.\ M.\ Maldacena,
{\sl Brane Dynamics From the Born-Infeld Action},
\npb{513}{1998}{198}, \hepth{9708147}.

\bibitem{Gibbons97}
G.\ W.\ Gibbons,
{\sl Born-Infeld Particles and Dirichlet p-branes}, 
\npb{514}{1998}{603}, \hepth{9709027}.

\bibitem{MN01}
D.\ Mateos and S.\ Ng, {\sl Supertubes Ending on D4-branes},
unpublished, 2001; to appear in S. Ng, {\sl Aspects of Branes in
String Theory}, PhD thesis, University of Cambridge, 2002.



\end{thebibliography}
\end{document}